\pdfoutput=1 %
\documentclass[sigconf, table, rgb, nonacm, natbib=false, screen]{acmart} 
\usepackage{lucas}
\usepackage{subcaption}
\usepackage{tikz}
\usetikzlibrary{backgrounds,fit,decorations.pathreplacing,calc,matrix,arrows.meta,positioning,decorations.pathreplacing,decorations.pathreplacing,calligraphy}
\usepackage{newtxmath}
\usepackage[textsize=tiny]{todonotes}
\usepackage{nicefrac}
\usepackage{csquotes}
\usepackage{mathtools}
\usepackage{enumitem}
\usepackage[detect-all=true]{siunitx}
\usepackage{url}
\usepackage{graphicx}
\usepackage{setspace}
\usepackage{multirow, makecell}
\usepackage{array}
\usepackage{marvosym}
\usepackage{xcolor}
\usepackage{verbatim}
\usepackage{nicematrix}
\usepackage{svg}

\usepackage{soul}
\usepackage{etoolbox}
\makeatletter
\patchcmd{\SOUL@ulunderline}{\dimen@}{\SOUL@dimen}{}{}
\patchcmd{\SOUL@ulunderline}{\dimen@}{\SOUL@dimen}{}{}
\patchcmd{\SOUL@ulunderline}{\dimen@}{\SOUL@dimen}{}{}
\newdimen\SOUL@dimen
\makeatother

\algblock{Input}{EndInput}
\algnotext{EndInput}
\algblock{Output}{EndOutput}
\algnotext{EndOutput}

\usepackage[style=ieee, maxnames=6, minnames=1, date=year, doi=false,isbn=false,backend=biber, sortcites, dashed=false]{biblatex}
\makeatletter
\let\oldtheequation\theequation
\renewcommand\tagform@[1]{\maketag@@@{\ignorespaces#1\unskip\@@italiccorr}}
\renewcommand\theequation{(\oldtheequation)}
\makeatother

\renewcommand*{\proofname}{Proof Sketch}

\renewcommand{\baselinestretch}{0.95}

\definecolor{myyellow}{RGB}{255, 253, 0}
\definecolor{myorange}{RGB}{238, 135, 51}
\definecolor{myblue}{RGB}{47, 112, 137}

\begin{document}
\title[Software Tools for Decoding Quantum Low-Density Parity Check Codes]{Software Tools for Decoding \\Quantum Low-Density Parity Check Codes\vspace*{-4mm}}
\author[Lucas Berent, Lukas Burgholzer, and Robert Wille]{Lucas Berent $^*$\hspace{2.0em}Lukas Burgholzer$^\ddagger$\hspace{2.0em}Robert Wille$^{*\dagger}$\hspace{2.0em}}
\affiliation{%
   \institution{$^*$Chair for Design Automation, Technical University of Munich, Germany}
}
\affiliation{%
   \institution{$^\ddagger$Institute for Integrated Circuits, Johannes Kepler University Linz, Austria}
}
\affiliation{%
  \institution{$^\dagger$Software Competence Center Hagenberg GmbH, Hagenberg, Austria}
}
\affiliation{lucas.berent@tum.de \hspace*{2.0em} lukas.burgholzer@jku.at \hspace*{2.0em} robert.wille@tum.de}

\begin{abstract}
\linespread{0.92}\selectfont
\emph{Quantum Error Correction} (QEC) is an essential field of research towards the realization of large-scale quantum computers. 
On the theoretical side, a lot of effort is put into designing error-correcting codes that protect quantum data from errors, which inevitably happen due to the noisy nature of quantum hardware and quantum bits (qubits). 
Protecting data with an error-correcting code necessitates means to recover the original data, given a potentially corrupted data set---a task referred to as \emph{decoding}. 
It is vital that decoding algorithms can recover error-free states in an efficient manner. 
While theoretical properties of recent QEC methods have been extensively studied, good techniques to analyze their performance in practically more relevant settings is still a widely unexplored area.
In this work, we propose a set of software tools that allows to numerically experiment with \mbox{so-called} 
\emph{Quantum Low-Density Parity Check codes} (QLDPC codes)---a broad class of codes, some of which have recently been shown to be asymptotically good.
Based on that, we provide an implementation of a general decoder for QLDPC codes.
On top of that, we propose an efficient heuristic decoder that tackles the runtime bottlenecks of the general QLDPC decoder while still maintaining comparable decoding performance.
These tools eventually allow to confirm theoretical results around QLDPC codes in a more practical setting and showcase the value of software tools (in addition to theoretical considerations) for investigating codes for practical applications.
The resulting tool, which is publicly available at \url{https://github.com/lucasberent/qecc} under the MIT license, is meant to provide a playground for the search for \emph{\enquote{practically good}} quantum codes.
\end{abstract}

\maketitle
\vspace*{-1mm}
\section{Introduction}
Current quantum computing research orbits around a central challenge, which is the physical realization of quantum computers. The main roadblock towards constructing universal, large-scale quantum computers is a fundamental problem that all quantum architectures suffer from: errors. Quantum systems are extremely susceptible to noise, which diminishes accuracy of computations and currently renders general quantum algorithms unusable in practice. Analogously to classical computing, \emph{Quantum Error Correction}~(QEC,~\cite{nielsen_quantum_2010}) evolves around designing methods that allow to protect quantum computers against noise and to reduce errors that inevitably happen in quantum systems in order to facilitate the realization of fault-tolerant quantum computers. 

\emph{Quantum Error-Correcting Codes} (QECCs)---designed to ward quantum systems by adding redundancy---are a main driver towards the goal of tackling noisy quantum hardware and achieving fault tolerance. This is due to results which state that with suitable QECCs it is possible to build arbitrarily large quantum computers in a fault-tolerant way. A problem with current QEC methods is that due to the added redundancy, the overall systems grow too large to actually build them. Recently, \emph{Quantum Low-Density Parity Check codes} (QLDPC codes)---a particular class of quantum error-correcting codes---have become the center of attention as they have good theoretical properties that promise applicability for large quantum systems. 

A central task in these endeavours is to efficiently recover a state that is encoded with a code and has potentially been corrupted---also referred to as \emph{decoding}.
Inefficient decoding leads to a bad overall performance of QEC techniques. 
On the one hand, decoders need to be fast so that the overall QEC performance is not diminished by the time it takes to decode.
On the other hand, they need to be able to correct as many errors as possible without introducing additional errors. 

Theoretical (asymptotic) properties of QLDPC codes, i.e., how much overhead they introduce and how many errors they can correct, have been studied thoroughly. 
This culminated in the construction of asymptotically good quantum codes~\cite{breuckmann_balanced_2021,panteleev_asymptotically_2022, leverrier_quantum_2022, dinur_good_2022}, i.e., codes whose good properties are preserved with increasing system size.
However, when viewed in a practical setting, i.e., with finite parameters (as opposed to asymptotic scaling), most codes are widely unexplored. Good theoretical properties do not necessarily imply that the codes also perform well for practical sizes (as also remarked in~\cite{panteleev_degenerate_2021}). 
Hence, investigations of codes with practical parameters are essential for future research towards fault-tolerant quantum computing.
In order to conduct such investigations around quantum codes and decoders, corresponding software tools are needed.

In this work, we propose such a set of software tools that allows to numerically experiment with the class of QLDPC codes that---to date---has mostly been explored theoretically.
Particularly focusing on the problem of decoding QLDPC codes,
we demonstrate how the resulting tool set can be employed to confirm theoretical results around QLDPC codes.
To this end, we implement a very general decoder that, in principle, can be applied to \emph{any} QLDPC code (based on the theoretical concepts provided in~\cite{delfosse_toward_2022}) and analyze its runtime and decoding performance.
On top of that, we propose a heuristic---based on ideas for decoding topological quantum codes---that tackles the inherent runtime bottleneck identified in the general decoder while still maintaining comparable decoding performance.
In an effort to continuously extend the amount of open source software tools for QEC, the resulting tool is made publicly available at \url{https://github.com/lucasberent/qecc}.

The rest of this paper is structured as follows:
\autoref{sec:background} covers the fundamental notions of QEC.
Then, \autoref{sec:motivation} reviews related work and the main motivation.
The general decoder provided as part of the proposed tool set is presented in~\autoref{sec:main}
and \autoref{sec:heuristic} discusses the implementation of the proposed heuristic.
Based on that,~\autoref{sec:numerical} summarizes the conducted numerical evaluations.
Finally, a short summary and an outlook on future directions are given in~\autoref{sec:conclusion}.

\section{Background}\label{sec:background}
To keep this work self-contained, this section covers the fundamental notions around ECCs and QEC.

\subsection{Error-Correcting Codes}
An \emph{Error-Correcting Code} (ECC) is a mechanism that adds redundancies to data to protect it from errors. A \emph{decoding} algorithm (decoder) tries to recover the original data from the (possibly erroneous) encoded one. 

In the following, data is simply viewed as binary vectors. Intuitively, a binary linear code $\calC{}$ of length $n$ can be seen as a set of vectors, called \emph{codewords}, which all fulfill the same set of \mbox{constraints (\emph{checks})}. Data vectors of length $k$ are encoded by assigning a codeword $x\in \calC{}$ (of length $n$) to each of them. Subsequently, when errors occur, the checks of $\calC{}$ are used to gain information on which error occurred in order to correct it accordingly. The checks compute the parity of a subset of elements $v_i$ of a vector $v$.

It is convenient to view a code $\calC{}$ as a bipartite graph 
$$\calT{}(\calC{}):=(V=V_Q\cup V_C,E),$$ called the \emph{Tanner graph} of $\calC{}$, where $V_C$ denotes the set of \emph{check} vertices and $V_Q$ the \emph{bit} (or data) vertices. Naturally, the set of bit vertices correspond to binary vectors and each check vertex checks if incident bits $v_i$ have even parity.

\begin{figure}[tb]
	\centering
	\centering
	\includesvg[width=0.4\linewidth,inkscapelatex=false]{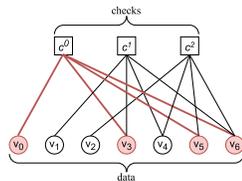}
\vspace*{-6mm}	
 \caption{Tanner graph of the Hamming code}
	\label{fig:hamming-tanner}
 \vspace*{-3mm}
\end{figure}

\begin{example}\label{ex:hamming-code}
	The length seven \emph{Hamming code} is an example of a binary linear code that encodes vectors of length four into vectors of length seven.
	\autoref{fig:hamming-tanner} depicts the Tanner graph of the Hamming code that has three check vertices (depicted as squares) and seven bit vertices (depicted as circles). For example, the left-most check~($c^0$) computes the parity of the bit vertices $v_0, v_3, v_5, v_6$.
\end{example}

Since the codewords are exactly the vectors that fulfill all checks, it is crucial to determine whether a given vector is a codeword or not in order to detect and correct errors. To check if a vector $v$ is a codeword, each bit node of $\calT{}(\calC{})$ is assigned an element $v_i\in \{0,1\}$ of $v$. Each check vertex $c^j$ computes the parity of its neighbours. If all checks are satisfied, i.e., the parity of the neighbours of each check is even, $v$ is a codeword.  

\begin{example}
Consider again~\autoref{ex:hamming-code} and let $x=(1,0,0,0,0,0,0)$ and $c^0$ be defined as above. Then, $x$ is not a codeword, since  
$$c^0\colon\; 1+0+0+0 \neq 0\; (\!\mathit{mod}\, 2).$$
\end{example}

Recall that if a vector $v$ is a codeword, all checks are satisfied. If any of the checks fails (computes an odd parity), it is indicated that an error $\varepsilon$ occurred. The checks that are not satisfied constitute the \emph{syndrome} of the error. To correct an error $\varepsilon$, an estimate vector $\tilde{\epsilon}$ that can be used to recover a codeword (an error-free state) has to be found. This process is called \emph{decoding}. Decoding algorithms are vital for error-correction, since the overall protection capability of a code depends on how well decoders can correct errors and how efficiently an estimate for a given syndrome can be found.

\subsection{Quantum Error Correction}
In classical computing, a fundamental unit of information is a bit,
which is protected by an ECC against bit
flip errors (flipping 1 to 0 and vice versa). The quantum analogue
of a bit is a \emph{qubit}~\cite{nielsen_quantum_2010}, which can assume arbitrary, complex linear
combinations of 0 and~1 (\emph{superposition}). The state $\ket{\psi}$ of a single qubit is
described as 
\mbox{$\ket{\psi}=\alpha_0\ket{0}+\alpha_1\ket{1}$}, where $\alpha_0,\alpha_1\in \C$ and \mbox{$\abs{\alpha_0}+\abs{\alpha_1}=1$}. 
It can be shown that errors on qubits are equivalent to combinations of bit flips and phase flips, where a bit flip is analogous to the classical case and
a phase flip on a qubit changes the sign of its $\ket{1}$ amplitude. 
Bit flips and phase flips correspond to applying
an $X$ or a $Z$ operator, respectively,
where $X$ and $Z$ are the well-known \emph{Pauli operators}.%

\begin{example}
The following equations showcase how $X$ and $Z$ operate on simple quantum states:
	\begin{align}
		&X\, \ket{0}=\ket{1},\; X\, \ket{1}=\ket{0} \\
		&Z\, \ket{0}=\ket{0},\; Z\, \ket{1}=-\ket{1}
	\end{align}
\end{example}

Since quantum errors are combinations of $X$ and $Z$ errors, it is natural to consider a combination of two classical codes, each protecting against one type of error. 
This is the main idea behind a broad class of codes
ubiquitous in quantum computing called \emph{Calderbank-Shor-Steane codes} (CSS codes,~\cite{calderbank_good_1996, steane_simple_1996}).

Analogously to a classical code, a CSS code $\calC{}$ can be represented as a Tanner graph $\calT{}(\calC{})$. The Tanner graph of a CSS code has two sets of check vertices, one for $Z$ errors and one for $X$ errors. 
\begin{figure}[tb]
	\centering
	\includesvg[width=0.38\linewidth]{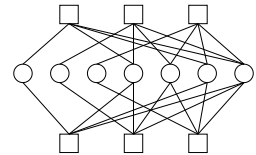}
\vspace*{-4mm}	
 \caption{Tanner graph of the Steane code}
	\label{fig:steane-tanner-both}
 \vspace*{-4mm}
\end{figure}

\begin{example}
	\autoref{fig:steane-tanner-both} depicts the Tanner graph of the \emph{Steane code}, which is a CSS code whose $X$ and $Z$ check components each correspond to a Hamming code (as illustrated before in \autoref{fig:hamming-tanner}).
\end{example}

The considered noise model assumes independent and identically distributed (i.i.d.) $X$ and $Z$ errors, hence the two components can be decoded independently and analogously to each other by finding an estimate for a given syndrome.
Thus, without loss of generality, we consider $X$ errors only in the following discussion. 
A \emph{Quantum Low-Density Parity Check} (QLDPC) code is a CSS code such that each bit is involved in a constant number of checks and each check involves a constant number of bits. This property renders QLDPC codes a promising candidate for practical applications, even for large code lengths.

\section{Motivation}\label{sec:motivation}
Recently, theoretical results around QLDPC codes have accumulated in several breakthroughs around asymptotically good quantum codes~\cite{breuckmann_balanced_2021, panteleev_asymptotically_2022, dinur_good_2022, leverrier_quantum_2022}.
These mostly theoretical results beg the question of practical applicability of such codes, e.g., whether they can be constructed and decoded efficiently for practical instance sizes.  
In fact, investigating the potential of these codes for practical applications has hardly been done yet and necessitates corresponding (software) tools. In this work, we take a step towards closing this gap. To this end, this section briefly reviews related work and remaining open problems before the remainder of this paper describes corresponding solutions.

\subsection{Related Work}\label{sec:rel-work}
Very recently, a generalized QLDPC decoder based on ideas of the \emph{Union-Find decoder} for topological codes was proposed~\cite{delfosse_toward_2022}. While the authors show theoretical decoding performance guarantees and conduct numerical simulations for a QLDPC code, no publicly available implementation is known and extensive numerical experiments on variants of the decoder and investigations around the practical runtime performance are still open.

Following the vast amount of research recently proposed in the domain of QEC, industrial research has also started to show interest in QEC and has acknowledged the importance of practical investigations. Most recent endeavors towards that include FlamingPy~\cite{tzitrin_fault-tolerant_2021} and Qiskit QEC (available at \href{https://github.com/qiskit-community/qiskit-qec}{github.com/qiskit-community/qiskit-qec}), which support toric codes but do not support the generalized QLDPC decoder. Furthermore, several tools~\cite{roffe_decoding_2020, roffe_ldpc_2022,higgott_pymatching_2021,gidney_stim_2021,higgott_improved_2022} have been used to numerically decode and evaluate quantum codes. Although some are publicly available,
most of them focus on surface codes and none include the recently proposed generalized QLDPC decoder.

\subsection{Considered Problem}
Since theoretical results around QLDPC codes have shown their promising properties, investigations of these codes in more practical settings (as opposed to asymptotic regimes) are of great importance. 
Central prerequisites for this are
software tools to facilitate the possibilities of research in this direction. 
In this work, we are proposing tools with a focus on the decoding problem for modern QLDPC codes. 
Decoding algorithms need to correct errors well and in a highly efficient manner, which naturally makes it a hard problem. In fact, decoders with sub-optimal performance have been identified as the main bottleneck of the QLDPC paradigm in QEC, as the general algorithm known from classical LDPC codes does not perform too well on QLDPCs~\cite{poulin_iterative_2008}. 

As a first step towards a comprehensive set of tools and methods, we focus on the decoding problem of QLDPC codes and provide implementations of a general QLDPC decoder in the form of an efficient, extendable, and openly available software tool set. 
On top of that and
to tackle inefficient runtime of current decoding algorithms, we additionally propose an algorithm to decode QLDPC codes with improved runtime performance. Both considerations are described in the following sections.

\section{Software Tools \\for Decoding QLDPC Codes}\label{sec:main}
While a decoder (based on ideas of decoding topological quantum codes) that, in principle, works for any QLDPC code has been proposed in~\cite{delfosse_toward_2022}, no software tools are available that implement this decoder and, hence, allow to perform broad numerical studies on QLDPC codes.
In the following, we describe our endeavours to change this situation.
First,~\autoref{sec:main-notation} settles the technical details that build the basis for the proposed software tools.
Based on that, \autoref{sec:original-uf} describes the implementation of the recently proposed general decoder for QLDPC codes.
Afterwards, \autoref{sec:discuss} discusses the main advantages and disadvantages of the resulting decoder.

\subsection{Technical Details}\label{sec:main-notation}

A CSS code $\calC{}$ is defined by two \emph{Parity Check Matrices} (PCMs) $H_X\in \F^{m\times n}_{2}$ and $H_Z\in \F^{l\times n}_{2}$, such that

\begin{equation}
	H_XH_{Z}^T=0.\label{eq:css-comm-cond}
\end{equation}

Such a code encodes $k=n-rank(H_X)-rank(H_Z)$ logical qubits into $n$ physical ones. 
Note that a PCM $H$ can be seen as an adjacency matrix of $\calT{}(\calC{})$, with rows corresponding to check vertices and columns to bit vertices. 
An entry $H_{ij}$ is set to one if and only if bit~$j$ occurs in the parity check defined by $c^i$. 

Concerning the possible errors that occur, we consider independent Pauli noise, where each data bit is affected by an error with a certain probability. For a CSS code of length $n$, an error is represented as a binary vector $\varepsilon=(x,z)\in \F^{n}_{2} \times \F^{n}_{2}$.

A CSS code has a set of \emph{stabilizers} which is a set of errors that have no effect on the encoded data. From~\autoref{eq:css-comm-cond} it follows that an error $x$ is a stablizer iff it is in the rowspace of $H_X$, i.e., a linear combination of $H_X$'s rows. 
The \emph{syndrome} of an error $x$, is $\sigma(x)=H_Zx \in \F^{l}_2$.

Given a syndrome $\sigma(x)$ of an error, the goal of the decoder is to find an estimate~$\tilde{x}$ that is equivalent to $x$ up to stabilizer, i.e., to find $\tilde{x}$ such that $x+\tilde{x}$ is in the rowspace of $H_X$. For simplicity, we assume that the syndrome can be inferred without any additional error being introduced during the process.
In the following, an $X$-error $x \in \F_{2}^{n}$ is identified with the qubit vertices $v\in V_Q$ it sets to one. Similarly, the syndrome $\sigma(x)$ of an error is identified with the set of check vertices $c^i \subseteq V_c$ triggered by $x$.

Tanner graphs are a central data structure for working with QLDPC codes.
To this end, some graph theoretic notions and functions are needed.
For any graph $G=(V,E)$ and a vertex $v\in V$, $N(v) \subseteq V$ denotes the neighbours of $v$, i.e., the vertices connected to $v$. 
The interior $Int(W)$ of a set of vertices $W$ contains vertices whose neighbours are also in $W$, i.e., $v\in Int(W)$ iff $N(v) \subseteq Int(W)$.

\begin{figure}[t]
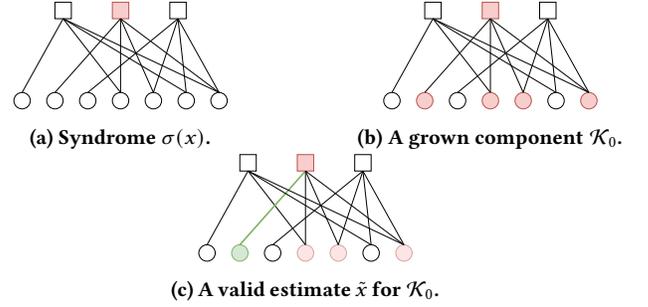

	\centering
	\begin{subfigure}[b]{0.20\textwidth}
		\centering
		\includesvg[width=0.8\textwidth]{tanndraw-Page-2.svg}
		\caption{Syndrome $\sigma(x)$.}
		\label{fig:steaneEx1}
	\end{subfigure}
	\hfill
	\begin{subfigure}[b]{0.20\textwidth}
		\centering
		\includesvg[width=0.8\textwidth]{tanndraw-Page-3.svg}
		\caption{A grown component $\calK{}_0$.}
		\label{fig:SteaneEx2}
	\end{subfigure}
	\begin{subfigure}[b]{0.20\textwidth}
		\centering
		\includesvg[width=0.8\textwidth]{tanndraw-Page-4.svg}
		\caption{A valid estimate $\tilde{x}$ for $\calK{}_0$.}
		\label{fig:SteaneEx3}
	\end{subfigure}\vspace*{-4mm}
	\caption{Example of UF decoding on the Steane code.}
	\label{fig:three graphs}
 \vspace*{-3mm}
\end{figure}

\begin{example}
The concepts described above are illustrated in \autoref{fig:three graphs} (the respective captions can be ignored for now).
To this end, \autoref{fig:steaneEx1} again shows the Tanner graph of the Steane code with the
check node $c^1$ marked red.
Its neighbours $N(c^1)$ are exactly the bit vertices also marked red in~\autoref{fig:SteaneEx2}. 
The interior of the marked vertices consists of $c^1$ together with the vertex highlighted in green as shown in \autoref{fig:SteaneEx3}.
\end{example}

\subsection{Implementation of a \\ General QLDPC Decoder}\label{sec:original-uf}
The technical details introduced above allow to describe the implementation of a decoder 
that can in principle decode any QLDPC code (based on the theoretical concepts provided in~\cite{delfosse_toward_2022}). 
This general QLDPC decoder uses ideas of the \emph{Union-Find} decoder for topological codes~\cite{delfosse_almost-linear_2021} and reduces the decoding problem of QLDPC codes to a combinatorial problem on the Tanner graph of the code. 
The general structure of the algorithm is sketched in~\autoref{alg:original}. 
In the following, an implementation of the algorithm is discussed.

Let $\calT{} = (V=V_Q \cup V_C, E)$ denote the Tanner graph of a QLDPC code.
Given a syndrome $\sigma(x) \subseteq V_C$ of some error, the decoder grows \emph{clusters} (sets of nodes) $\calK{}_i \subseteq V$ around syndrome nodes until all clusters are \emph{valid} (Line~\autoref{lin:general-grow}). A cluster $\calK{}_i$ is valid if an estimate $\tilde{x}_i$ can be found that covers the syndrome occurring in $\calK{}_i$. Formally, a cluster $\calK{}_i$ is defined to be valid iff there exists a set of nodes $\tilde{x}_i \subseteq V_Q \cap Int(\calK{}_i)$ such that $\sigma(\tilde{x}_i) = \sigma(x) \cap \calK{}_i$.
When all grown clusters $\calK{}_i$ are valid, an overall estimate $\tilde{x}$ whose syndrome matches $\sigma(x)$ is computed by looking for local corrections $\tilde{x}_i$ inside each $\calK{}_i$ (Line~\autoref{lin:general-subcorr}).

\begin{algorithm}[t]
\caption{Generalized QLDPC Decoder~\cite{delfosse_toward_2022}}\label{alg:original}
\begin{algorithmic}[1]
	\State Input: $\sigma(x) \subseteq V_C$ of some error $x$.
	\State Output: Estimate $\tilde{x}$ s.t. $\sigma(\tilde{x})=\sigma(x)$
	
	\State Initialize $\calK{} = \sigma(x)$
	
	\While {$\calK{}$ contains invalid clusters}\label{lin:general-while}
	\State Grow $\calK{}_i$ by adding $N(v), \, \forall v \in \calK{}_i$\label{lin:general-grow}
	\EndWhile
	\For {$\calK{}_i \subseteq \calK{}$}
	\State Find valid correction $\tilde{x}_i \subseteq \text{Int}(\calK{}_i)$ with $\sigma(\tilde{x}_i) = \sigma(x) \cap \calK{}_i$\label{lin:general-subcorr}
	\EndFor
\end{algorithmic}
\end{algorithm}

\begin{example}
Consider again
\autoref{fig:three graphs}, which suffices to illustrates the main idea of the generalized QLDPC decoder. Assume that check $c^1$ fails.
First (\autoref{fig:steaneEx1}), the syndrome $\sigma(x)=c^1$ is initialized as the first (and only) cluster $\calK_0$. Then, $\calK{}_0$ is grown~(\autoref{fig:SteaneEx2}). The nodes of $\calK{}_0$ are marked. The decoder can find a correction $\tilde{x}=(0,1,0,0,0,0,0)$, as there is a qubit node in $Int(\calK{}_0)$ covering the syndrome~(\autoref{fig:SteaneEx3}).
\end{example}

\subsection{Discussion}\label{sec:discuss}
The advantages of the algorithm are clearly its generality which implies that it can be applied for any QLDPC code. 
Furthermore, the algorithm has a rather simple formulation as it is formalized as a graph problem. 
Moreover, the authors of~\cite{delfosse_toward_2022} proved correction guarantees for several classes of codes.
More specifically, the decoder can successfully decode all sufficiently low-weight errors for hyperbolic QLDPC codes of arbitrary dimension~\cite{guth_quantum_2014,londe_golden_2017}, toric codes~\cite{kitaev_fault-tolerant_2003}, and locally testable codes~\cite{leverrier_towards_2022,hastings_decoding_2014,leverrier_quantum_2015}. 
On a practical note, the proposed implementation only requires the formulation of codes in terms of parity-check matrices, which is very general and therefore ensures compatibility with a broad range of codes and easy extendability.
The main bottlenecks of this decoder are the procedures for checking validity and finding corrections which, essentially, can be reduced to Gaussian elimination~\cite{delfosse_toward_2022}. 
Overall, \autoref{alg:original} thus has worst-case runtime complexity $O(n^4)$.

\section{Union-Find Decoding Heuristic}\label{sec:heuristic}
In order to tackle the runtime complexity of the decoder discussed above, in this section, we propose decoding heuristic based on ideas borrowed from a decoder for topological quantum codes~\cite{delfosse_almost-linear_2021} that improves the runtime of the general QLDPC decoder by a quadratic factor and show how it can be implemented on top of the proposed software package.

\subsection{General Ideas}
In order to achieve an efficient runtime performance, a dedicated data structure, called Union-Find (UF, also called disjoint-set data structure,~\cite{cormen_introduction_2009}), is employed to represent clusters of nodes.
This data structure accelerates two of the main routines in the general decoder considered above: determining the cluster a node belongs to (the \emph{Find} function) and efficiently merging two separate clusters (the \emph{Union} function).

The Union-Find data structure consists of a list of disjoint sets of vertices (of $\calT{}$). 
Each disjoint set is represented as a tree (\emph{UF tree}) rooted at a dedicated node $r_i$. 
The overall data structure can then be seen as list of root nodes---each identifying a UF tree. 
Because of the tree-like representation, the operations of determining which UF tree a given node belongs to (\emph{Find}), and merging two disjoint UF trees (\emph{Union}) can be computed efficiently. 

It is important to emphasize that the proposed algorithm is a heuristic, since there is no (theoretical) guarantee it can correct a certain number of errors for general QLDPC codes. 
However, the runtime performance is considerably reduced by a factor of $O(n^2)$.
Therefore, the proposed heuristic
presents an interesting step in the direction of highly efficient decoding algorithms and allows to trade-off decoding performance for runtime.

\begin{algorithm}[tb]
\caption{Union-Find Decoding Heuristic}\label{algorithm:improved}
\begin{algorithmic}[1]
	\State Input: $\sigma(x) \subseteq V_C$ 
	\State Output: Estimate $\tilde{x} \subseteq V_Q$ s.t. $\sigma(\tilde{x})=\sigma(x)$
	
	\State Initialize $\calK{} = \sigma(x)$
	\State Init $\calA = \emptyset$
	\While {$\calK{} \neq \emptyset$}
	\State For all $\calK{}_i \in \calK{}$ grow $\calK{}_i$ by its neighbourhood\label{lin:alg-growth}
	\State Use \emph{Union} to merge clusters grown together\label{lin:alg-union}
	\State Replace each cluster representative $u_i$ by its root using \emph{Find}\label{lin:alg-root-replace}
	\State Update boundary lists \label{lin:alg-boundarylist-update}
	\State Add valid clusters to $\calA{}$ \label{lin:alg-add-valid-to-sol}
	\EndWhile
	\Procedure{Erasure procedure}{$\calA$}\label{lin:alg-improved-erasure-decoding} \Comment{Local Decoding}
	\State Compute $Int(\calK{}_i)$
	\State Take a bit node $v_j \in Int(\calK{}_i)$
	\State Add $v_j$ to the estimate $\tilde{x}_i$ for $\calK{}_i$
	\State Remove checks $c^k$ adjacent to $v_j$ by removing $N(c^k)$
	\EndProcedure
\end{algorithmic}
\end{algorithm}

\subsection{Implementation} 
Using the idea of the generalized QLDPC decoder and the UF data-structure as sketched above, \autoref{algorithm:improved} sketches the implementation of the proposed approach. Based on the UF data structure, clusters grown in the Tanner graph $\calT$ are represented as UF trees. The algorithm maintains a list \mbox{$\calK{}=\set{r_0,\dots, r_\ell}$} of root nodes $r_i$ corresponding to the grown clusters $\calK{}_i$.
As the algorithm grows clusters in $\calT{}$ by adding the neighbouring vertices (in $\calT{}$) of boundary nodes to a cluster (Line \autoref{lin:alg-growth}), it may happen that two clusters are grown together,~i.e., share a common node. The \emph{Find} function is used to determine which cluster a node belongs to and consecutively \emph{Union} is used to efficiently compute the union of two UF trees, i.e., merge two clusters into one (Line~\ref{lin:alg-union}). Once all clusters are valid, a procedure to compute an estimate $\tilde{x}_i$ is computed for each cluster $\calK{}_i$ (Line~\autoref{lin:alg-improved-erasure-decoding}).

To maintain low runtime, some further implementation details have to be considered.
The algorithm grows clusters in $\calT{}$ until all are valid, i.e., until each cluster $\calK{}_i$, $Int(\calK{}_i)$ contains a possible correction. The validity check determines for each cluster whether there exists neighbours of check nodes that are not in the boundary. 
To avoid recomputation of the boundary in each step (which would contribute to a quadratic factor in the runtime), a precomputed list of \emph{boundary} vertices is stored at the root node of each UF tree. The boundary list of a cluster $\calK{}_i$ contains nodes of the cluster that share an edge with a node not in the cluster. Thus a book keeping step is needed that recomputes the boundary list for each $\calK{}_i$ after the growth step (Line \autoref{lin:alg-boundarylist-update}). 

Moreover, similar to~\cite{delfosse_almost-linear_2021}, it is necessary to replace each vertex representing a cluster by its root node in $\calK{}$. Merging of clusters might cause that this produces duplicates. Thus, a lookup table indicating which root is present in $\calK{}$ is added. To avoid duplicate root nodes, the decoder can simply check whether a root is already present, and if so not add it again to $\calK{}$.

The main cost of growing clusters and is dominated by the the Find and Union functions. Using weighted Union (appending smaller to larger clusters) and path compression (when calling Find, appending all nodes encountered directly as children of the root node) leads to an almost linear worst-case time complexity since both functions have a worst-case complexity of $O(n \alpha(n))$ where $\alpha$ is the inverse Ackermann function~\cite{tarjan_efficiency_1975}. Checking validity and finding estimates for each cluster requires quadratic worst-case runtime, since for each cluster a valid estimate is computed iteratively from the nodes inside the cluster. 
Overall the worst-case runtime is in $O(n^2)$---a considerable improvement compared to the general QLDPC decoder. 

\section{Numerical Evaluations}\label{sec:numerical}
The ideas presented in the previous sections have been implemented in C++ with easy-to-use Python bindings as an interface.
The resulting implementation is available at \url{https://github.com/lucasberent/qecc}.
Based on that, we investigate the runtime scaling as well as the decoder performance of both presented decoders.
Additionally, we study the impact of different variants of growth on the performance of the decoding performance.
As mentioned above, in all of our investigations, we assume independent and identically distributed Pauli noise, perfect syndrome measurements, and focus on $X$ errors only. 
To construct the codes, we used a publicly available tool~\cite{roffe_ldpc_2022}. The codes and all numerical data is made publicly available along with the source code. All numerical simulations were conducted on a machine equipped with a 16 core Intel Xenon(R) W-1370P $3.6$GHz processor and $128$GiB RAM, running Ubuntu $20.04$ LTS.

\subsection{Runtime Performance}
\begin{figure}
    \centering
    \includesvg[width=0.95\linewidth]{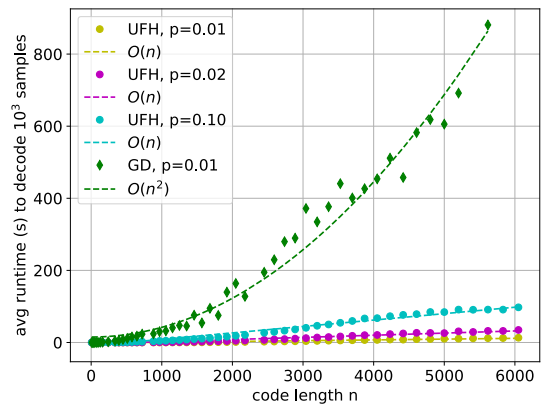}
    \vspace*{-4mm}
    \caption{Average runtime of the general QLDPC decoder and the heuristic to decode $10^3$ samples for toric codes with increasing length $n$.}
    \label{fig:rt}
\end{figure}

In a first series of evaluations, the runtime of the general QLDPC decoder and the proposed heuristic was investigated. To this end, we ran Monte Carlo simulations for decoding toric codes of increasing sizes under different levels of physical noise (physical X-error rate, per).
\autoref{fig:rt} depicts the obtained average runtimes for the general QLDPC decoder (GD) and the heuristic (UFH). Note that each data point corresponds to the average runtime to decode $10^3$ samples. 
The results confirm the discussion in \autoref{sec:discuss} on the bottleneck of the general QLDPC decoder---the use of Gaussian elimination. This is the dominating factor here since the degree of the Tanner graph is small. The experimental data shows a scaling in $O(n^2)$ for our implementation, the considered code, and noise model.

The data clearly confirms the drastic performance benefit of the heuristic. For small code sizes the runtime behaves almost constantly. For code sizes $n$ larger than roughly $10^3$, linear scaling can be observed while maintaining low runtimes even for large code sizes. For instance, for code sizes of $n=3000$ with $p=0.01$, the heuristic takes several seconds to decode $10^3$ samples, while the general QLDPC decoder needs several minutes.

\subsection{Decoding Performance}
In addition to the runtime behaviour, the decoding performance of the general QLDPC decoder and the proposed heuristic was studied on a medium-sized QLDPC code. 
More specifically, we conducted simulations in which errors with increasing physical error rate were sampled and the fraction of failed runs with respect to the code dimension (the so-called \emph{Word-Error Rate}, WER) was investigated for a $[\![1024,18]\!]$ lifted product code~\cite{panteleev_degenerate_2021}.

We used a simulation procedure that is constituted of repeated runs of the following steps for a code with parameters $[\![n,k]\!]$:
\begin{enumerate}
\item Sample a Pauli error $x \in \F^{n}_2$.
\item Compute the syndrome $\sigma(x)$.
\item Apply the decoder to get $\tilde{x}$.
\item Compute $x' = x + \tilde{x}$.
\item Check if $x'$ is a stabilizer. If yes, then return success otherwise return failure.
\end{enumerate}

The fraction of failed runs is called the block error rate, $P_L$. 
In order to factor out the number of encoded qubits $k$, the \emph{Word-Error Rate} (WER) $P_W$ defined as
\[P_W=1-(1-P_L)^{1/k}.\]
was used to evaluate the decoding performance.
For small logical error rates it holds that $P_W \approx P_L/k$. The growth procedure in Line~\ref{lin:alg-growth} of \autoref{algorithm:improved} offers some degrees of freedom for different variants of how this growth is realized.
In our evaluation we compared several different variants of growth. The first one (AG), where all clusters are grown in each growth step (also valid ones) is motivated by the original formulation~\cite{delfosse_toward_2022} and underpinned by theoretical results that guarantee correctability bounds for the general QLDPC decoder. The other two variants grow only a single cluster in each growth step, either the smallest one (SSG) or a random one (SRG). Single smallest cluster growth has been demonstrated to be beneficial for topological codes~\cite{delfosse_almost-linear_2021}.

\begin{figure}
    \centering
    \includesvg[width=0.95\linewidth]{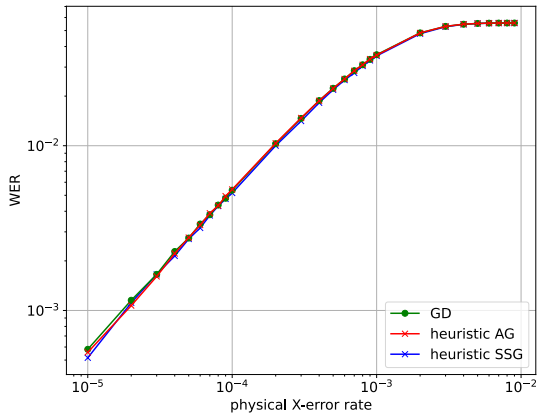}    \vspace*{-3mm}
    \caption{Decoding performance of the proposed heuristic using different variants of growth and the general QLDPC decoder on a $[\![1024,18]\!]$ lifted product code.}
    \label{fig:dp}
\end{figure}

The results of the investigations around the decoding performance are summarized in~\autoref{fig:dp}. Every data point was obtained by $10^5$ samples. Since the growth variants of the proposed heuristic where a single cluster is grown in each step (a random one or the smallest one) perform very similarly, only the single smallest cluster growth is depicted. 

In summary, the overall WER of both the general decoder and the proposed heuristic is rather high for the considered physical error rates. This matches results from~\cite{delfosse_toward_2022}, where the authors demonstrate that the general QLDPC decoder only outperforms a belief-propagation decoder for very low physical error-rates.
Interestingly, all growth variants of the proposed heuristic perform very similarly. Furthermore, the heuristic performs nearly equivalently to the general QLDPC decoder in the considered scenario.

\section{Conclusion}\label{sec:conclusion}
Motivated by several recent \emph{theoretical} breakthrough results around QLDPC codes, this work provides a set of software tools for investigations in more \emph{practical} settings. The central aspect studied in this work is the decoding of QLDPC codes. 
The proposed solution encompasses an open-source implementation of a general QLDPC decoder
and allows to conduct numerical evaluations on a large variety of codes to investigate the runtime as well as the decoding performance of this algorithm. 
On top of that, we proposed an algorithm that aims to address the runtime bottlenecks of the general decoder while still maintaining comparable decoding performance.
Both solutions have been evaluated with regard to their runtime as well as decoding performance---confirming theoretical results and showcasing promising future directions.

In the future, it would be interesting to improve the decoding performance---for both the original decoder and the proposed heuristic---for instance by considering more sophisticated growth procedures. Especially ones that consider edge weights and hence \emph{weighted growth}, or a combination with a pre-decoder. Moreover, taking more realistic noise models into account is a crucial next step. Furthermore, it would be desirable to prove decoding performance bounds for families of codes for the proposed heuristic analytically.

\section*{Acknowledgements}
The authors would like to thank Richard Kueng and James Wootton for fruitful discussions in the early stages of this project and for insightful comments on a first version of this paper, respectively.\vspace{0.5mm}
This work received funding from the European Research Council (ERC) under the European Union’s Horizon 2020 research and innovation program (grant agreement No. 101001318), was part of the Munich Quantum Valley, which is supported by the Bavarian state government with funds from the Hightech Agenda Bayern Plus, and has been supported by the BMWK on the basis of a decision by the German Bundestag through project QuaST, as well as by the BMK, BMDW, and the State of Upper Austria in the frame of the COMET program (managed by the FFG).
\printbibliography
\end{document}